\documentclass[letterpaper, 10 pt, conference]{ieeeconf}  % Comment this line out if you need a4paper

\IEEEoverridecommandlockouts                    % This command is only needed if 
\usepackage{amsmath} % assumes amsmath package installed
\usepackage{amssymb}  % assumes amsmath package installed
\usepackage{algorithm}
\usepackage[noend]{algpseudocode}
\usepackage{color}
\usepackage{graphicx}
\usepackage{booktabs}
\usepackage{enumerate}
\usepackage{setspace}

\title{\LARGE \bf Learning a Better Control Barrier Function} 

\author{Bolun Dai$^{1}$, Prashanth Krishnamurthy$^{1}$, Farshad Khorrami$^{1}$
\thanks{$^{1}$Control/Robotics Research Laboratory, Electrical~\&~Computer Engineering Department, Tandon School of Engineering, New York University, Brooklyn, NY, 11201
{\tt\small bd1555@nyu.edu, prashanth.krishnamurthy@nyu.edu, khorrami@nyu.edu.}
}
}

\newcommand{\F}{\mathbf{F}}
\newcommand{\G}{\mathbf{G}}

\newcommand{\bR}{\mathbb{R}}

\newcommand{\bu}{\mathbf{u}}

\newcommand{\x}{\mathbf{x}}
\newcommand{\dx}{\dot{\mathbf{x}}}
\newcommand{\y}{\mathbf{y}}

\begin{document}

\maketitle
\thispagestyle{empty}
\pagestyle{empty}

\maketitle
\setstretch{1.1}
\begin{abstract}
Control barrier functions (CBFs) are widely used in safety-critical controllers. However, constructing a valid CBF is challenging, especially under nonlinear or non-convex constraints and for high relative degree systems. Meanwhile, finding a conservative CBF that only recovers a portion of the true safe set is usually possible. In this work, starting from a ``conservative" handcrafted CBF (HCBF), we develop a method to find a CBF that recovers a reasonably larger portion of the safe set. Since the learned CBF controller is not guaranteed to be safe during training iterations, we use a model predictive controller (MPC) to ensure safety during training. Using the collected trajectory data containing safe and unsafe interactions, we train a neural network to estimate the difference between the HCBF and a CBF that recovers a closer solution to the true safe set. With our proposed approach, we can generate safe controllers that are less conservative and computationally more efficient. We validate our approach on two systems: a second-order integrator and a ball-on-beam.
\end{abstract}

% Using a different approach, we can safely generate solutions within the true safe set by incorporating the hard constraints into an optimal control problem, e.g., model predictive control (MPC). 

% Model predictive control (MPC) is another widely used approach to generate safe control actions by incorporating the hard constraints into an optimal control problem.

% Nevertheless, such an approach is usually computationally expensive and may not lend itself to real-time implementations. We propose to combine the two methods. 

% During training, we utilize model predictive control (MPC) to collect safe trajectory data.
\section{Introduction}
% Safe Control
Safety is one of the most critical aspects when designing a controller~\cite{DBLP:conf/amcc/DaiKPK21}~\cite{DBLP:journals/corr/abs-2107-07931}. With self-driving cars~\cite{DBLP:conf/iros/SrinivasanDCV20} and robot manipulators~\cite{DBLP:conf/iros/SaverianoL19} in our daily lives, generating safe controllers has become increasingly important. Work has been done in generating safe trajectories using trajectory optimization~\cite{DBLP:conf/iros/HowellJM19}. However, safe controllers must be able to react to changing situations quickly, which requires small computational costs. Trajectory optimization techniques are computationally expensive, especially for nonlinear systems with long trajectories~\cite{DBLP:journals/siamrev/Kelly17}. To reduce the computational cost, work has been done to learn a safe controller. However, learning-based controllers usually require unsafe interactions~\cite{DBLP:conf/amcc/DaiKPK21}. Recently, work has been done to ensure safety during learning~\cite{DBLP:journals/ral/ThananjeyanBNLS21} by switching between a safe backup and an exploration controller. However, this approach generates jerky motions, making it less desirable for real-world applications.

% CBF-based Safe Control
CBFs provides a simple yet effective way to synthesize controllers for safety-critical applications in a minimally invasive fashion~\cite{DBLP:conf/eucc/AmesCENST19}. However, as the complexity of the constraints increases, especially for nonlinear or non-convex constraints and nonlinear systems with a high relative degree, it becomes increasingly difficult to construct an HCBF, e.g., dynamic walking on stepping stones~\cite{DBLP:conf/cdc/NguyenHGAS16}. However, finding an HCBF that only recovers a portion of the safe set is usually possible~\cite{DBLP:conf/cdc/ChoiLSTH21}. Instead of finding CBFs by hand, work has been done on learning CBFs using human demonstrations~\cite{DBLP:conf/iros/SaverianoL19} of state trajectories along the safe set boundary, which are not always available. CBFs can also be learned from a dataset of safe and unsafe trajectories~\cite{DBLP:conf/cdc/RobeyHLZDTM20}. In many applications, unsafe trajectories might be costly or even dangerous to obtain, e.g., autonomous driving. Instead of relying on external resources, CBFs can also be synthesized online using onboard sensors~\cite{DBLP:conf/iros/SrinivasanDCV20}. This approach has the benefit of using only safe training data. However, it might not generalize well outside object avoidance tasks since not all safe set boundaries can be obtained through sensor measurements. In addition to finding a good CBF, work has been done on finding a looser CBF constraint under a given CBF~\cite{DBLP:journals/corr/abs-2201-01347}. However, this does not mitigate the issue of having conservative CBFs.

% MPC
Another way to ensure safety is to incorporate the safety requirements as constraints of an optimization problem, as in MPC~\cite{845037}. MPC has been widely applied, e.g., autonomous driving~\cite{kong2015kinematic} and bipedal locomotion~\cite{DBLP:conf/iros/HerdtPW10}. However, compared to the solution time of a CBF quadratic program (CBF-QP), solving MPC problems is usually time-consuming, which reduces applicability to real-time control. Thus, a multi-time scale control framework is typically used. MPC generates a reference trajectory at a lower frequency while a reactive controller runs at a higher frequency. Another issue with MPC is that the constraints only affect the control within the time horizon. This can be alleviated by increasing the time horizon with the cost of also increasing solution time.

% Contribution
In this paper, we propose an algorithmic approach to combine CBF and MPC-based approaches in learning a CBF that recovers a larger portion of the true safe set. The main contribution of this paper is twofold: (1) we develop a method to estimate the true CBF starting from an HCBF; (2) we show the effectiveness of the proposed method through simulation studies. The remainder of this paper is structured as follows. In Section II, the CBF and MPC approaches are briefly summarized. In Section III, the problem formulation for this paper is given. In Section IV, the proposed method is presented. In Section V, we show the efficacy of our approach using simulation studies on two systems: second-order integrator and  ball-on-beam. Finally, we conclude the paper with a summary and a discussion on future directions.
\section{Preliminaries}
\subsection{Control Barrier Function Based Safe Control}
This section summarizes the approach of achieving certifiable safety critical control using CBFs. For a detailed review please refer to~\cite{DBLP:conf/eucc/AmesCENST19}. Consider a control affine system
\begin{equation}
    \dx = \F(\x) + \G(\x)\bu
    \label{eq:system_dynamics}
\end{equation}
with the states being denoted by $\x\in\bR^{n}$, the control inputs by $\bu\in\bR^{m}$, the drift by $\F: \bR^n\rightarrow\bR^n$, and the control influence by $\G: \bR^n\rightarrow\bR^{n\times m}$. Additionally, we have a safe set $\mathcal{C}$ defined as the superlevel set of a smooth function $\mathbf{h}: \bR^n \rightarrow \bR$, i.e.,
\begin{equation}
\label{eq:cbf_condition}
    \mathcal{C} = \{\x\ \mid\ \mathbf{h}(\x) \geq 0\}.
\end{equation}
We can synthesize safe controllers by finding controls that satisfy the constraint
\begin{equation}
    \frac{\partial\mathbf{h}(\x)}{\partial\x}\dx \geq -\alpha(\mathbf{h}(\x))
    \label{eq:cbf_qp_constraint}
\end{equation}
where $\alpha(\cdot)$ is a class $\mathcal{K}_\infty$ function. The inequality~\eqref{eq:cbf_qp_constraint} is called the CBF constraint. Starting from a performance controller $\pi_\mathrm{perf}: \bR^n \rightarrow \bR^m$, we can use the CBF constraint as a safety filter and solve for a minimally invasive safe control. This approach is called CBF-QP. The performance controller first provides a reference control action $\bu_\mathrm{ref} \sim \pi_\mathrm{perf}(\x)$. Note that this performance controller does not need to generate control actions that ensure the forward invariance of the safe set. Then, the CBF-QP is constructed as
\begin{align}
\label{eq:cbf-qp}
    \min_\bu\ \ \ \ & \|\bu_\mathrm{ref} - \bu\|^2\\
    \mathrm{subject\ to}\ \ \ \ & L_{\mathbf{F}}\mathbf{h}(\x) + L_{\mathbf{G}}\mathbf{h}(\x)\bu \geq -\alpha(\mathbf{h}(\x))\nonumber
\end{align}
where $L_{\beta}(\cdot)$ represents the Lie derivative of $(\cdot)$ with respect to $\beta$. The solution to~\eqref{eq:cbf-qp}, $\bu_\mathrm{safe}$, is the safe control action.

\subsection{Model Predictive Control}
MPC is a widely used framework for solving finite-horizon optimal control problems. Define the instantaneous cost at time $t$ as $\ell(\x(t), \bu(t), t)$. The MPC objective is to minimize the sum of the instantaneous costs over a predefined time horizon $T$ subject to constraints, which can be written as
\begin{align}
\label{eq:mpc_formulation}
    \min_{\bu}\ &\ \int_{t_0}^{t_0 + T}\ell(\x(t), \bu(t), t)dt\\
    \mathrm{subject\ to}\ &\ \dx(t) = \F(\x(t)) + \G(\x(t))\bu(t)\nonumber\\
    &\ \mathbf{c}(\x(t)) \leq \mathbf{b}\nonumber
\end{align}
with $\mathbf{c}: \bR^n\rightarrow\bR^r$ being a vector of functions of the state and $\mathbf{b}\in\bR^r$ representing the element-wise upper bound of $\mathbf{c}(\cdot)$. Note that control saturation can also be incorporated, but it is not included in~\eqref{eq:mpc_formulation} for brevity. The solution to the MPC problem gives the optimal control actions over the time interval $[t_0,\ t_0+T]$. MPC provides a robust framework for generating control actions by iterating this process. One major drawback for MPC is that with the increase in horizon length, the solution time would also grow, making it difficult to deploy directly to real-time control tasks. In general, CBF-QPs can be solved much faster than MPCs.
\section{Problem Formulation}
In this work, we consider systems in the form of~\eqref{eq:system_dynamics}. We have the following assumptions: (a) we have access to a performance controller $\pi_\mathrm{perf}(\x)$ that enables a system in the form of~\eqref{eq:system_dynamics} to achieve a certain task, e.g., target reaching; (b) we have access to an accurate model of the true system dynamics (this is a commonly used assumption in previous works on CBFs~\cite{DBLP:conf/cdc/RobeyHLZDTM20}~\cite{DBLP:journals/corr/abs-2201-01347}); (c) the safe set $\mathcal{C}$ is defined as the largest set in $\{\x\mid\mathbf{c}(\x) \leq \mathbf{b}\}$ where $\exists\bu$ that renders it forward invariant; (d) we do not have access to an accurate CBF; (e) we have access to the function $\alpha(\cdot)$. 

We define the CBF that recovers the safe set $\mathcal{C}$ as $\mathbf{h}(\x)$, which satisfies~\eqref{eq:cbf_condition}. Utilizing $\mathbf{h}(\x)$, we can formulate a CBF-QP that gives us the safe control action $\bu_\mathrm{safe}$. However, in general, it is difficult to find a CBF that recovers the true safe set $\mathcal{C}$~\cite{DBLP:conf/cdc/ChoiLSTH21}. But it is usually viable to find a CBF $\widehat{\mathbf{h}}(\x)$ that recovers a subset of the true safe set, i.e., 
\begin{equation}
    \widehat{\mathcal{C}} = \{\x\mid\widehat{\mathbf{h}}(\x) \geq 0\} \subseteq \mathcal{C}.
\end{equation}
We refer to such a CBF $\widehat{\mathbf{h}}$ that we can find analytically as a ``handcrafted" CBF. As discussed above, handcrafted CBFs are often conservative in practice since they provide an underestimate of the true safe region. Without loss of generality, we can assume the relationship
\begin{equation}
    \mathbf{h}(\x) = \widehat{\mathbf{h}}(\x) + \Delta\mathbf{h}(\x).
\end{equation}
In this work, we aim to find an algorithmic approach to generate an estimate of $\Delta{\mathbf{h}}(\x)$. Define the safe set recovered by the estimated CBF as
\begin{equation}
    \widetilde{\mathcal{C}} = \{\x \mid \widehat{\mathbf{h}}(\x)+\Delta\widehat{\mathbf{h}}(\x) \geq 0\}
\end{equation}
where $\Delta\widehat{\mathbf{h}}(\x)$ is an estimation of $\Delta{\mathbf{h}}(\x)$. The following relationship should hold 
\begin{equation}
\label{eq:set_relationship}
    \widehat{\mathcal{C}} \subseteq \widetilde{\mathcal{C}} \approx \mathcal{C}.
\end{equation}
% When recovering the true safe set is not relevant to achieving the given task, the expanded safe set $\widetilde{\mathcal{C}}$ should ensure the task is better achieved, e.g., having a lower cost.
\section{Method}
In this section, we detail our proposed approach of learning an estimate of the true CBF starting from an HCBF. We separate the discussion into two sections: CBF learning and data collection.

\begin{figure*}[t!]
    \centering
    \includegraphics[width=\textwidth]{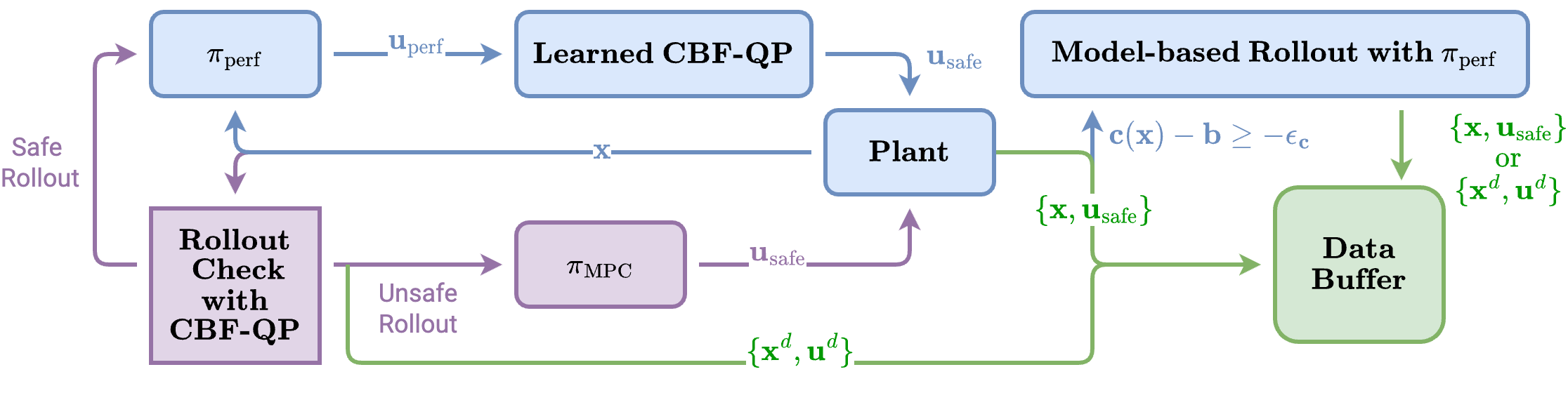}
    \caption{This figure illustrates the data collection process. The blue arrows represents the control loop that is performed at each timestep. The purple arrows represents the control loop that is performed every $r_\mathbf{CBFQP}$ timesteps. The green arrows represents storing the collected data.}
    \label{fig:BetterCBF}
\end{figure*}

\subsection{CBF Learning}
\label{sec:learning_alg}
In this section, we discuss how $\Delta\widehat{\mathbf{h}}(\x)$ is learned such that
\begin{equation}
    \widetilde{\mathbf{h}}(\x) = \widehat{\mathbf{h}}(\x) + \Delta\widehat{\mathbf{h}}(\x)
\end{equation}
is a good estimation of the true CBF $\mathbf{h}(\x)$. To learn $\Delta\widehat{\mathbf{h}}(\x)$, we use a neural network parameterized by $\theta$ (see
Section~\ref{sec:ddn} for details). The learned CBF is denoted as
\begin{equation}
    \widetilde{\mathbf{h}}(\x\mid\theta) = \widehat{\mathbf{h}}(\x) + \Delta\widehat{\mathbf{h}}(\x\mid\theta).
\end{equation}
In the remainder of this section, the dependency of $\widetilde{\mathbf{h}}$ and $\Delta\widehat{\mathbf{h}}$ on $\theta$ will be omitted for brevity. The condition a valid CBF should satisfy is shown in~\eqref{eq:cbf_condition}. In addition, to obtain safe control actions, the partial derivative of $\widetilde{\mathbf{h}}(\x)$ w.r.t. the state should also satisfy
\begin{equation}
    L_{\mathbf{F}}\widetilde{\mathbf{h}}(\x) + L_{\mathbf{G}}\widetilde{\mathbf{h}}(\x)\bu + \alpha(\widetilde{\mathbf{h}}(\x)) \geq 0\ \ \ \ \forall\x\in\mathcal{C}
    \label{eq:cbf_constraint_condition}
\end{equation}
with $\bu$ being a safe action.
% CBF Safe Loss
Although~\eqref{eq:cbf_condition} is the sole requirement for CBF values within the safe set, we can add additional constraints to the CBF value to guide the learning process. We force the CBF value for safe states to increase the further away the states are from the constraint boundary. Note that the CBF values for states on the constraint boundary should be less than or equal to zero to ensure the relationship in~\eqref{eq:set_relationship}. The above considerations yield
\begin{equation}
    \widehat{\mathbf{h}}(\mathbf{x}_i) + \Delta{\widehat{\mathbf{h}}(\mathbf{x}_i)} \geq \mathbf{d}_+(\x_i)\ \ \ \ \forall\mathbf{x}\in\mathcal{C}
    \label{eq:cbf_condition_safe}
\end{equation}
with $\mathbf{d}_+:\mathbb{R}^n\rightarrow\mathbb{R}$ being a function computing the distance between the current state and the constraint boundary. For examples regarding the choice of $\mathbf{d}_+(\x)$ please refer to Section~\ref{sec:simulation}. The loss function for safe data is defined as
\begin{equation}
    \mathcal{L}_\mathbf{h} = \frac{1}{N_\mathrm{safe}}\sum_{i=1}^{N_\mathrm{safe}}\max\Big(0, -\widehat{\mathbf{h}}(\mathbf{x}_i) - \Delta{\widehat{\mathbf{h}}(\mathbf{x}_i)} + \mathbf{d}_+(\x_i)\Big)
    \label{eq:cbf_loss}
\end{equation}
where $N_\mathrm{safe}$ is the size of the safe dataset. It can be seen that $\mathcal{L}_\mathbf{h}$ is minimized when~\eqref{eq:cbf_condition_safe} is satisfied for all $\mathbf{x}_i$'s. 
% CBF Unsafe Loss
Like safe interactions, even though~\eqref{eq:cbf_condition} is the only condition required for a valid CBF, the learning process would greatly benefit from stronger conditions. We force the CBF value for unsafe data points to decrease as it moves further into the interior of the unsafe set. Using another function $\mathbf{d}_{-}(\x)\in\mathbb{R}_{-}$, we can write the condition unsafe data points should satisfy as
\begin{equation}
    \widehat{\mathbf{h}}(\mathbf{x}_i) + \Delta{\widehat{\mathbf{h}}(\mathbf{x}_i)} \leq \mathbf{d}_{-}(\x)\ \ \ \ \forall\mathbf{x}\in\backslash\mathcal{C}.
    \label{eq:cbf_condition_unsafe}
\end{equation}
Rewriting the condition in~\eqref{eq:cbf_condition_unsafe} into a loss function, we have
\begin{equation}
    \mathcal{L}_\mathbf{d} = \frac{1}{N_d}\sum_{i=1}^{N_d}\max\Big(0,\  \widehat{\mathbf{h}}(\mathbf{x}_i^d) + \Delta{\widehat{\mathbf{h}}(\mathbf{x}_i^d)} - \mathbf{d}_{-}(\x_i^d) \Big)
\end{equation}
where $N_d$ is the size of the unsafe dataset. The states $\mathbf{x}_i^d$ are sampled from the unsafe (dangerous) dataset.

% CBF Constraint Loss
To synthesize safe control actions using the learned CBF, we would also require it to satisfy the CBF constraints~\eqref{eq:cbf_constraint_condition}. Using the same logic in constructing~\eqref{eq:cbf_loss}, we can write a loss function that is minimized when all $\x_i$'s satisfies~\eqref{eq:cbf_constraint_condition}: 
\begin{align}
    \label{eq:cbf_constraint_loss}
    \mathcal{L}_\mathbf{\nabla h} =&\ \frac{1}{N_\mathrm{safe}}\sum_{i=1}^{N_\mathrm{safe}}\max\Big(0,\\
    &\ -L_{\mathbf{F}}\widetilde{\mathbf{h}}(\x_i) - L_{\mathbf{G}}\widetilde{\mathbf{h}}(\x_i)\bu_i - \alpha(\widetilde{\mathbf{h}}(\x_i))\Big)\nonumber.
\end{align}

% dCBF Size Loss
When there are more safe data than unsafe data (the generation of unsafe data will be discussed in Section~\ref{sec:data_collection}), we need to ensure that the learned CBF does not simply output a positive value for any $\mathbf{x}$. This can be achieved by regulating the change introduced by $\Delta{\widehat{\mathbf{h}}(\mathbf{x})}$. The logic behind this choice is that for unsafe states, the HCBF $\widehat{\mathbf{h}}(\mathbf{x})$ already gives them a negative value. By regulating the change introduced by $\Delta{\widehat{\mathbf{h}}(\mathbf{x}\mid\theta)}$, we aim to only make the CBF values for the safe states positive while keeping the unsafe states negative. To achieve this, we construct the loss function
\begin{equation}
    \mathcal{L}_{\Delta\mathbf{h}} = \frac{1}{N_\mathrm{safe}}\sum_{i=1}^{N_\mathrm{safe}}{\Delta\widehat{\mathbf{h}}^2(\mathbf{x}_i)}.
\end{equation}
However, when the amount of safe and unsafe data is balanced, this loss can be removed.
% Final Loss
We then have the final loss function as
\begin{equation}
    \mathcal{L} = \mathcal{L}_\mathbf{h} + \lambda_1\mathcal{L}_\mathbf{d} + \mathcal{L}_\mathbf{\nabla h} + \lambda_2\mathcal{L}_{\Delta\mathbf{h}}
    \label{eq:loss_function}
\end{equation}
where $\lambda_1, \lambda_2\in\mathbb{R}_+$ weights the importance of $\mathcal{L}_\mathbf{d}$ and $\mathcal{L}_{\Delta\mathbf{h}}$, respectively. Since estimating part of the unsafe set as safe is much more disastrous than estimating part of the safe set as unsafe, $\lambda_1$, if not zero (when there are no unsafe data), is usually set to be larger than one. The choice of $\lambda_2$ is task-dependent. Some example settings can be seen in Section~\ref{sec:simulation}.
% Neural Network
We represent the safe and unsafe dataset as $\mathcal{X} = \{\mathbf{x}_i\}_{i=1}^{N_{\mathrm{safe}}}$ and $\mathcal{X}_\mathbf{d} = \{\mathbf{x}_i^d\}_{i=1}^{N_d}$, where $\mathcal{X}$ is the safe dataset and $\mathcal{X}_\mathbf{d}$ is the unsafe data set. The collection process of the datasets will be described in Section~\ref{sec:data_collection}.

\begin{algorithm}[t]
\caption{Algorithm for Learning a Better CBF}
\begin{algorithmic}[1]
    \State {\bf Given}: $\mathbf{F}(\mathbf{x})$, $\mathbf{G}(\mathbf{x})$: state dynamics,  $\widehat{\mathbf{h}}(\mathbf{x})$: handcrafted CBF, $\alpha(\cdot)$: class $\mathcal{K}_\infty$ function, $\pi_\mathrm{perf}$, $\pi_\mathrm{MPC}$: performance controller and MPC controller, $N$: episode length, $L$: number of epochs, $M$: batch size, $\theta_1$: initialized neural network weights, $\lambda_1$, $\lambda_2$: loss weights;
    \For {$l = 1$ to $L$}
        \State Sample initial state $\x_\mathrm{init}$;
        \For {$n = 1$ to $N$}
            \State $\mathbf{u}_\mathrm{perf} = \pi_\mathrm{perf}(\mathbf{x}_n)$;
            \State Run CBF-QP using $\widetilde{\mathbf{h}}(\x\mid\theta)$ and get $\mathbf{u}_\mathrm{safe}$;
            \If {$n\ \%\ r_\mathrm{CBFQP} = 0$}
                \State Perform model-based rollout using CBF-QP
                \If {unsafe state encountered}
                    \State $\mathbf{u}_\mathrm{safe} = \pi_\mathrm{MPC}(\x)$
                \EndIf
            \EndIf
            \If {$\mathbf{c}(\x) - \mathbf{b} \geq -\epsilon_\mathbf{c}$}
                \State Perform model-based rollout using $\pi_\mathrm{perf}$
            \EndIf
            \State Apply $\mathbf{u}_\mathrm{safe}$ and store safe and unsafe data;
        \EndFor
        \State Update $\theta_l$ using $\mathcal{L}$ and obtain $\theta_{l+1}$;
    \EndFor
\end{algorithmic}
\label{alg:algo}
\end{algorithm}

\subsection{Data Collection}
\label{sec:data_collection}
As mentioned in Section~\ref{sec:learning_alg}, learning $\widetilde{\mathbf{h}}(\x\mid\theta)$ requires both safe and unsafe data. However, collecting unsafe data may be costly in many cases, e.g., autonomous driving. Instead of finding an unsafe controller, we perform model-based rollouts (forward simulation for computation of trajectory samples) to generate ``fake" data. We use a CBF-QP controller with the learned CBF during training to collect safe interactions. Note that unless $\Delta{\widehat{\mathbf{h}}(\mathbf{x}\mid\theta)}$ is fully trained, there is no guarantee that the learned CBF corresponds to a safe set. To deal with this issue, the controller performs a rollout using the CBF-QP controller every $r_\mathbf{CBFQP}$ steps. If the rollout generates unsafe data, the data is recorded, and an MPC controller is invoked to generate safe control actions. Using this approach, we can obtain safe and unsafe data without subjecting the real system to unsafe conditions. In addition to the model-based rollout mentioned above, a model-based rollout can also be performed when the system is close to the constraint boundary. To check if the current state is close to the constraint boundary, we define $\epsilon_\mathbf{c}\in\mathbb{R}_+$, and whenever at least one of the entries of $\mathbf{c}(\x) - \mathbf{b}$ is larger than $-\epsilon_\mathbf{c}$  a model-based rollout is performed. The value of $\epsilon_\mathbf{c}$ is task-dependent. When performing the rollout, $\pi_\mathrm{perf}$ is used to generate control actions, which is generally an unsafe controller. A diagram of the data collection process is given in Figure~\ref{fig:BetterCBF}. Psuedo-code for the learning and data collection process can be found in Alg.~\ref{alg:algo}. Since training is performed offline, the computational costs of MPC can be mitigated by using simulation environments or more computational resources than during inference. 

% \begin{subequations}
% \begin{align}
%     \label{eq:cbf_safe_condition}
%     \widetilde{\mathbf{h}}(\x) &> 0\ \ \ \ \forall\x\in\mathrm{Int}\ \mathcal{C}\\
%     \label{eq:cbf_unsafe_condition}
%     \widetilde{\mathbf{h}}(\x) &< 0\ \ \ \ \forall\x\in\backslash\mathcal{C}.
% \end{align}
% \end{subequations}
\section{Simulation}
\label{sec:simulation}
In this section, we demonstrate our proposed approach on two tasks. The first task, second-order integrator target reaching with velocity constraint, is a toy example where the learned CBF can be plotted and analyzed. The second task, ball-on-beam balancing with angle constraint, demonstrates our approach on a nonlinear system. It also shows the robustness of our proposed method by showing the CBF constraint can be altered after training to produce controllers with varying aggressiveness when approaching the constraint boundary. All of the training is performed using PyTorch. All of the MPC controllers use a time horizon of 20 time steps.

\subsection{Deep Differential Network}
\label{sec:ddn}
When utilizing the learned CBF in the simulation studies, a CBF-QP is solved to find the safe control. Therefore, we need to compute $\partial\Delta\widehat{\mathbf{h}}(\x\mid\theta) / \partial\x$. We use deep differential networks~\cite{DBLP:conf/iclr/LutterRP19} to compute the Jacobians in closed form with machine precision. At each layer, the partial derivative of the output $\y^\prime\in\mathbb{R}^{n_o\times1}$ w.r.t. the input $\y\in\mathbb{R}^{n_i\times1}$ is
\begin{equation}
    \frac{\partial\y^\prime}{\partial\y} = \mathrm{diag}(d\mathbf{g}(\mathbf{a}))\mathbf{W}
\end{equation}
where $\mathbf{W}\in\mathbb{R}^{n_o\times n_i}$ represents the weights of that layer, $\mathbf{g}: \mathbb{R}^{n_o\times1}\rightarrow\mathbb{R}^{n_o\times1}$ represents the activation function, $d\mathbf{g}(\cdot)$ represents the derivative of the activation function, $\mathbf{a} = \mathbf{W}\y + \mathbf{bias}$, $n_i$ is the size of the input, and $n_o$ is the size of the output. In all our experiments, we use a network of three layers. The output size of each layer is $[128, 128, 1]$. All of the activation functions are Softplus.

\subsection{Second-Order Integrator Target Reaching}
\begin{figure}[t!]
    \centering
    \includegraphics[width=0.49\textwidth]{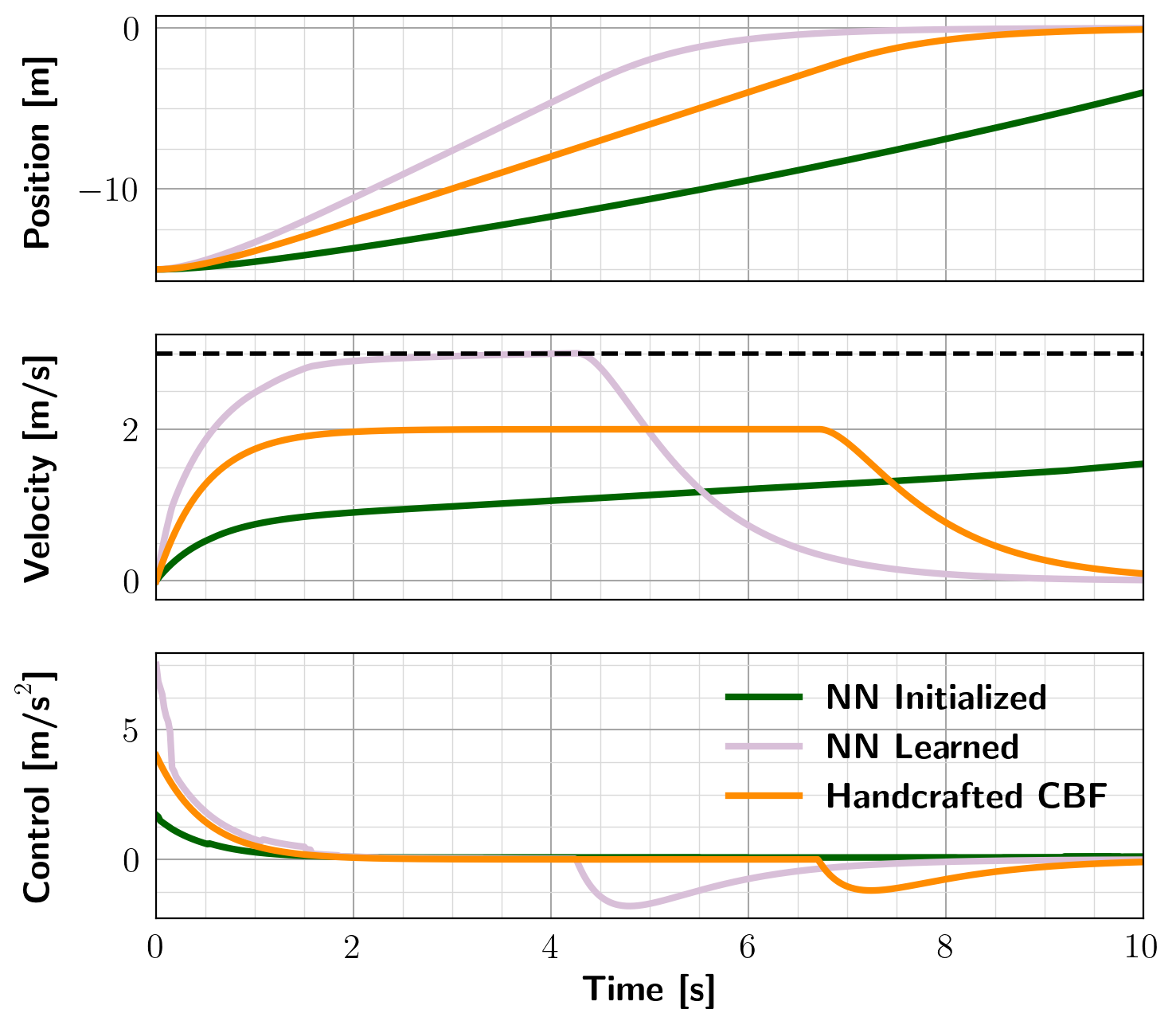}
    \caption{This figure shows the state and control trajectories of the CBF-QP controller using different CBFs. The NN Initialized trajectory is obtained using the CBF $\widehat{\mathbf{h}}(\x) + \Delta\widehat{\mathbf{h}}(\x\mid\theta_1)$, the NN Learned is obtained using the CBF $\widehat{\mathbf{h}}(\x) + \Delta\widehat{\mathbf{h}}(\x\mid\theta_L)$, and the Handcrafted CBF trajectory is obtained using the CBF $\widehat{\mathbf{h}}(\x)$. The dashed line in the velocity plot corresponds to $\dot{x} = 3$, which represents the velocity constraint.}
    \label{fig:integrator2dexp}
\end{figure}
\begin{figure}[t!]
    \centering
    \includegraphics[width=0.49\textwidth]{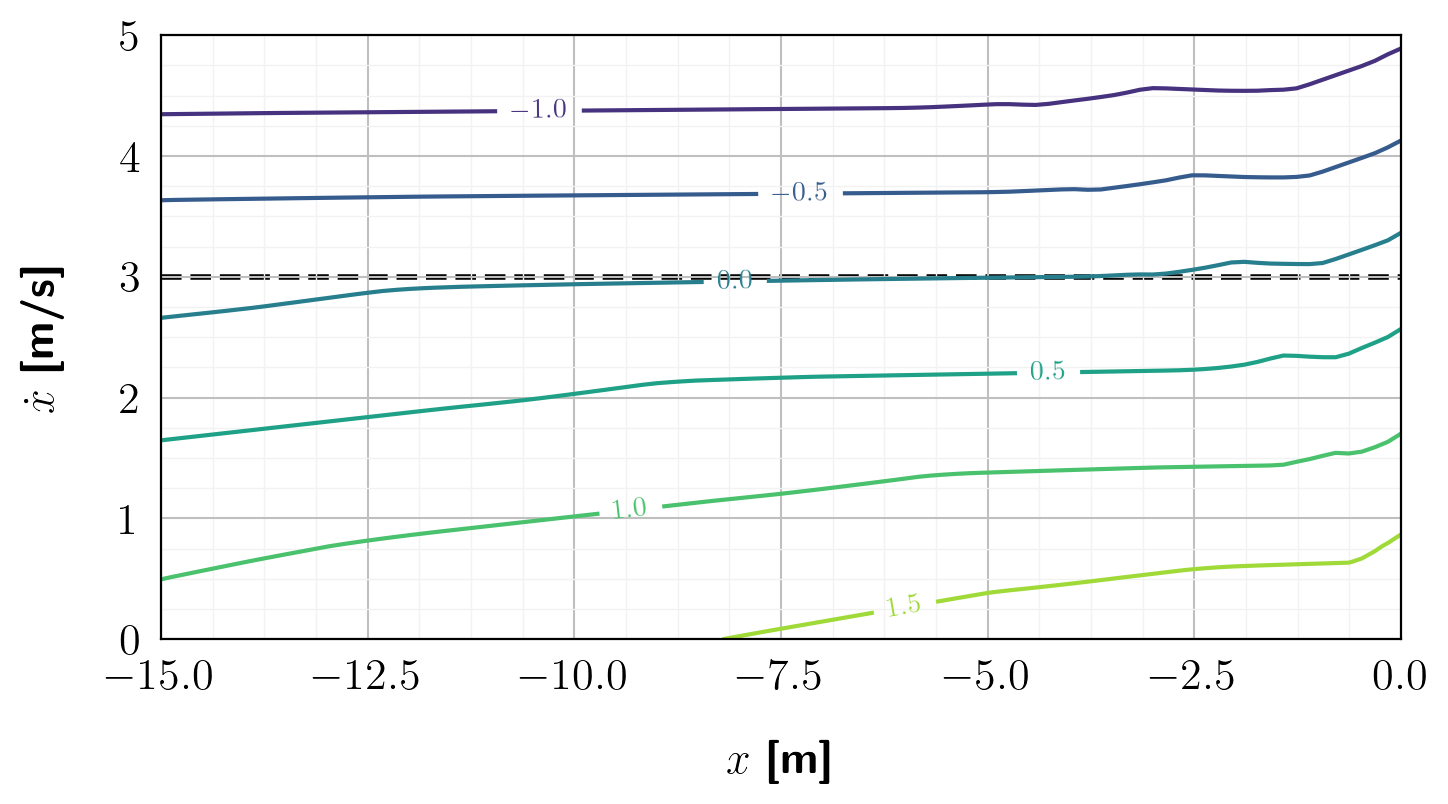}
    \caption{This figure shows the contour of the learned CBF. The black dashed line denotes the zero level line of the true CBF, the region above the dashed line will be negative for the true CBF and positive below.}
    \label{fig:integrator2d_learnedCBF}
\end{figure}
The second-order integrator has the system dynamics of
\begin{equation}
\label{eq:2dintegrator_dynamics}
    \begin{bmatrix}
        \dot{x}\\
        \ddot{x}
    \end{bmatrix} = \begin{bmatrix}
        0 & 1\\
        0 & 0
    \end{bmatrix}\begin{bmatrix}
        x\\
        \dot{x}
    \end{bmatrix} + \begin{bmatrix}
        0\\
        1
    \end{bmatrix}\bu
\end{equation}
with $x$ denoting the position, $\dot{x}$ the velocity, and $\ddot{x}$ the acceleration. The dynamics~\eqref{eq:2dintegrator_dynamics} can also be written as $\dot{\x} = \mathbf{A}\x + \mathbf{B}\bu$, with $\x\in\bR^{2}$, $\mathbf{A}\in\bR^{2\times2}$, $\mathbf{B}\in\bR^{2\times1}$, and $\bu\in\bR$. The velocity constraint is $\dot{x} \leq 3.0$. We construct an HCBF $\widehat{\mathbf{h}}(\x) = 2 - \dot{x}$, which corresponds to the velocity constraint $\dot{x} \leq 2.0$. The task is to start from $\x_\mathrm{init} = [x_\mathrm{init},\ 0]^T$ and reach the origin. The performance controller is an LQR controller with the weights $\mathbf{Q} = \mathrm{diag}(10, 10)$ and $\mathbf{R} = [1]$. The MPC formulation is
\begin{align}
    \min_{\bu_i\mathrm{'s}}\ &\ \sum_{i=1}^{T}{\Big(\frac{1}{2}\x_i^T\mathbf{Q}\x_i + \frac{1}{2}\bu_i^T\mathbf{R}\bu_i\Big)}\\
    \mathrm{subject\ to}\ &\ \x_0 = \x,\ \dot\x_{i} = \mathbf{A}\x_i + \mathbf{B}\bu_i,\ \dot{x}_i \leq 3.0\nonumber
\end{align}
where $T$ is the time horizon. For this example, a discrete time step of 0.02s is used. We set $\lambda_1 = 0$ and $\lambda_2 = 1$. The CBF is trained for 100 epochs with a learning rate of $10^{-3}$ and the initial position $x_\mathrm{init}$ is randomly sampled from $[-15, -5]$. The distance functions are
\begin{equation}
    \mathbf{d}_{+}(\x) = \mathbf{d}_{-}(\x) = 3 - \dot{x}.   
\end{equation}
We pick the class $\mathcal{K}_\infty$ function $\alpha(x) = \gamma x$ for~\eqref{eq:cbf_constraint_condition}, with $\gamma = 5.0$ during training. Since this is a toy example, we can also construct a CBF for the true constraint $\mathbf{h}(\x) = 3 - \dot{x}$ and compare it with the learned CBF. Under the aforementioned setting, using Algorithm~\ref{alg:algo}, the state trajectory of the learned CBF-QP controller is shown in Figure~\ref{fig:integrator2dexp}, with $\x_\mathrm{init} = -15$. During training, both CBF-QP-based rollouts and model-based rollouts are performed. The contour plot of the learned CBF is shown in Figure~\ref{fig:integrator2d_learnedCBF}. From Figure~\ref{fig:integrator2dexp}, we can see that the learned CBF can recover the original velocity constraint. The HCBF curve underestimates the velocity constraints, while the NN initialized curve depends on the initialization of the neural network weights. From Figure~\ref{fig:integrator2d_learnedCBF}, we can see that the zero level set of the learned CBF is almost the same as the zero level set of the true CBF. There are inaccuracies in estimating where the zero level line should be when $x$ is close to $-15$ and $0$. This is due to having no data with $\dot{x}$ close to $3$m/s at the beginning and the end of the trajectory.

\subsection{Ball-on-beam Balancing}
\begin{figure}[t!]
    \centering
    \includegraphics[width=0.49\textwidth]{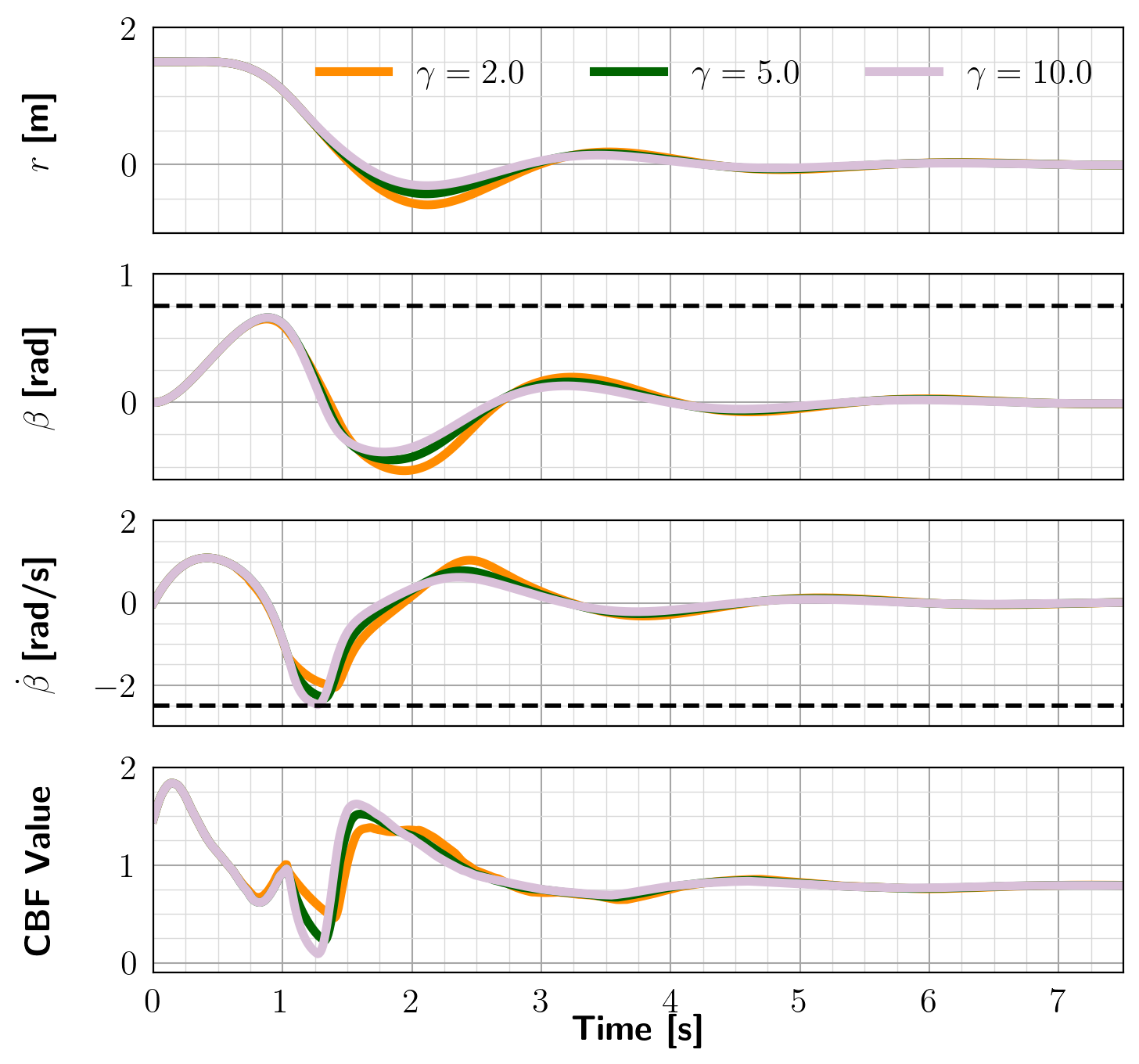}
    \caption{This figure illustrates the motion and CBF values under the learned CBF controller for the ball-on-beam simulations. The upper three plots illustrate the motion generated by the learned CBF under different values of $\alpha(\mathbf{h}(\x)) = \gamma\mathbf{h}(\x)$ in the CBF-QP constraint. The dashed black line represents the constraints $\beta \leq 0.75$ and $\dot{\beta} \geq -2.5$. The lowest plot illustrates the CBF values along the trajectories.}
    \label{fig:ballonbeam_exp}
\end{figure}
The ball-on-beam system has the dynamics
\begin{equation}
\label{eq:ballonbeam_dynamics}
    \begin{bmatrix}
        \dot{r}\\
        \dot{\beta}\\
        \ddot{r}\\
        \ddot{\beta}
    \end{bmatrix} = \begin{bmatrix}
        \dot{r}\\
        \dot{\beta}\\
        \displaystyle\frac{5}{7}(r\dot{\beta}^2 - g\sin\beta)\\
        \displaystyle-\frac{2mr\dot{r}\dot{\beta} + mgr\cos\beta}{I_\mathrm{beam} + mr^2}
    \end{bmatrix} + \begin{bmatrix}
        0\\
        0\\
        0\\
        \displaystyle\frac{1}{I_\mathrm{beam} + mr^2}
    \end{bmatrix}\mathbf{u}
\end{equation}
with $r$ being the position of the ball along the beam, $\beta$ being the angle of the beam, $m$ the mass of the ball, $g$ the gravitational acceleration, $I_\mathrm{beam}$ the moment of inertia of the beam, and $\mathbf{u}$ the input torque applied to the beam. The objective is to move the ball from $r = r_\mathrm{init}$ to $r = 0$ considering the beam angle and angular velocity constraints $\beta \leq 0.75$ and $\dot{\beta} \geq -2.5$. The initial state is $[r_\mathrm{init},\ 0,\ 0,\ 0]^T$, the target state is $[0,\ 0,\ 0,\ 0]^T$.

It is difficult to construct a conservative HCBF that combines these two constraints. Instead, we start from the HCBF
\begin{equation}
\label{eq:ball-on-beam_hcbf}
    \widehat{\mathbf{h}}(\x) = -\dot{\beta} + \gamma_0(\bar{\beta} - \beta),\quad\beta\leq\bar{\beta} < 0.75
\end{equation}
which corresponds to a tighter beam angle constraint so that the motion generated under the HCBF would also not violate the angular velocity constraint. For this example, we choose $\bar{\beta} = 0.5$. Starting from~\eqref{eq:ball-on-beam_hcbf}, we learn a CBF that can recover the safe set described by the original constraints. The performance controller is an LQR controller using the linearized dynamics with the cost matrices as $\mathbf{Q} = \mathrm{diag}(10, 1, 1, 1)$ and $\mathbf{R} = 1$. The nonlinear MPC (NMPC) is defined as
\begingroup
\allowdisplaybreaks
\begin{align}
    \min_{\bu_i\mathrm{'s}}\ &\ \sum_{i=1}^{T}{\Big(\frac{1}{2}\x_i^T\mathbf{Q}\x_i + \frac{1}{2}\bu_i^T\mathbf{R}\bu_i\Big)}\\
    \mathrm{subject\ to}\ &~\eqref{eq:system_dynamics},\ \x_0 = \x,\ \beta \leq 0.75,\ \dot{\beta} \geq -2.5.\nonumber
\end{align}
\endgroup
For this example, a discrete time step of 0.01s is used. The NMPC is solved using \texttt{do-mpc}~\cite{LUCIA201751}. We set $\lambda_1 = 2$, $\lambda_2 = 0$. The class $\mathcal{K}_\infty$ function $\alpha(x) = \gamma x$, with $\gamma = 2.0$ during training. The distance functions are
\begingroup
\allowdisplaybreaks
\begin{subequations}
\begin{align}
    \mathbf{d}_{+}(\x) &= \min(0.75 - \beta, \dot{\beta} + 2.5)\\
    \mathbf{d}_{-}(\x) &= \max(0.75 - \beta, \dot{\beta} + 2.5).
\end{align}
\end{subequations}
\endgroup
The CBF is trained for 1000 epochs with a learning rate of $10^{-4}$. During training, only CBF-QP-based rollouts are performed. The motions generated using the learned CBF is shown in Figure~\ref{fig:ballonbeam_exp}. To display the robustness of the learned CBF, we tested the learned CBF using different values of $\gamma$. The generated motions can also be seen in Figure~\ref{fig:ballonbeam_exp}. The trajectories generated by different $\gamma$ values are all safe trajectories. The learned CBF values along the trajectories are shown in Figure~\ref{fig:ballonbeam_exp}. As $\gamma$ increases, the smallest CBF value along the trajectory gets closer to zero. This example illustrates how to apply our proposed approach to nonlinear systems and that the learned CBF constraint can be tweaked after training.
\section{Conclusion}
In this paper, we propose a learning-based approach to estimate the CBF that recovers a larger portion of the true safe set starting from an HCBF. We designed an algorithmic data collection procedure that ensures safety. Additionally, unsafe data are generated synthetically during training using model-based rollouts. We tested our proposed approach on two different systems and tasks. Potential future directions include applying to more complex systems and incorporating inaccurate or learned dynamics.

\vspace*{-0.05in}
\bibliographystyle{IEEEtran}
\bibliography{IEEEabrv, refs.bib}
\end{document}